\documentclass[preprint]{jpsj2}

\def\runtitle{Instructions for the Preparation of Manuscript}
\def\runauthor{Yuki \textsc{Fuseya},
Hiroshi \textsc{Kohno} and Kazumasa \textsc{Miyake}}

\title{%
Realization of Odd-Frequency $p$-Wave Spin-Singlet 
Superconductivity Coexisting with Antiferromagnetic Order near
Quantum Critical Point
}

\author{%
Yuki \textsc{Fuseya},
Hiroshi \textsc{Kohno} and
Kazumasa \textsc{Miyake}
}

\inst{%
Division of Materials Physics,
Department of Physical Science, Graduate School of Engineering Science, \\
Osaka University, Toyonaka, Osaka 560-8531, Japan
}

\recdate{\today}

\abst{It is shown that the $p$-wave singlet superconductivity, whose gap
function is odd both in momentum and in Matsubara frequency, is realized
prevailing over the $d$-wave singlet state very close to
the antiferromagnetic quantum critical point (QCP) both in the paramagnetic
and in antiferromagnetic sides. Off the QCP in the paramagnetic phase, however,
the $d$-wave singlet superconductivity with line-nodes is realized
as \textit{conventional} anisotropic superconductivity.
This $p$-wave singlet superconductivity is an essentially gapless state,
i.e., there is no gap in the quasiparticle spectrum everywhere
on the Fermi surface due to its 
odd-frequency gap. These features can give a qualitative
but nice understanding of the anomalous behaviors of 
NQR relaxation rate on CeCu$_2$Si$_2$ or CeRhIn$_5$ 
where the antiferromagnetism and superconductivity
coexist on a microscopic level. 
}

\kword{$p$-wave singlet superconductivity, gapless superconductivity,
coexistence of antiferromagnetism and superconductivity,
CeCu$_2$Si$_2$, CeRhIn$_5$
}
\newcommand{\Tc}{T_{\textrm{c}}}
\newcommand{\siml}{\stackrel{<}{_\sim}}
\newcommand{\simg}{\stackrel{>}{_\sim}}

\begin{document}
\sloppy
\maketitle

\section{Introduction}
	The discovery of the superconductivity in CeCu$_2$Si$_2$ \cite{Steglich}
	two decades ago
	triggered off a breakthrough in the field of superconductivity.
	Nowadays, we seem to have reached the consensus that 
	the anisotropic spin singlet pairing, e.g. ``$d$-wave
	singlet" pairing, is realized in the superconducting (SC) states 
	close to the antiferromagnetic (AF) states due to the AF spin fluctuations.
	\cite{MSV,Scalapino,Monthoux,Ueda}
	Recent developments on CeCu$_2$Si$_2$, however, suggest that 
	it still disguises much various physics.
	\cite{Onishi,Ishida,YKawasaki1,YKawasaki2,Koda,SO5,Kitaoka1,Yuan}
	 Polycrystalline sample of Ce$_{0.99}$Cu$_{2.02}$Si$_2$ shows a peak
	 in the nuclear spin lattice relaxation rate $1/T_1 $ at
	 SC transition temperature $\Tc \sim 0.65$K
	 and  $1/T_1 \propto T$ behavior at $T \leq \Tc$, 
	 which indicates essentially gapless superconductivity
	 to be realized at $T < \Tc$ at ambient pressure.
	At pressures $P \simg 0.1$GPa, however, the $1/T_1 \propto
	T^3 $ behavior, indicating line-node gap, is observed.
	\cite{Ishida,YKawasaki1,SO5}
	Since the Ge substitution for Si is considered to expand the lattice
	constant, giving negative pressure, 1\% Ge substituted compound 
	CeCu$_2$(Si$_{0.99}$Ge$_{0.01}$)$_2$ exhibits
	AF ordering at $T< T_{\textrm N}=0.75$K, followed by the onset
	of superconductivity at $T=\Tc =0.5$K at ambient pressure.
	The observation of single component of NQR signal, showing an 
	appearance of the internal field throughout the sample,
	excludes the possibility of phase segregation between the SC
	and the AF phases; namely, SC and AF coexist on a microscopic level.	
	In this SC states, $1/T_1$ does not show any significant reduction
	below $\Tc$,  but shows essentially the same behavior
	as the normal Fermi liquid state.
	At $P=0.85$GPa, on the other hand, it does not exhibit the AF ordering
	and the line-node SC gap evidenced from a clear behavior of 
	$1/T_1 \propto T^3$
	\cite{YKawasaki2}.

	Interestingly, the recently discovered pressure induced superconductor,
	CeRhIn$_5$\cite{Heeger} of Ce-based heavy fermion, 
	shows features quite similar to those of CeCu$_2$Si$_2$,
	while its crystal structure is rather two-dimensional 
	compared to CeCu$_2$Si$_2$.
	At ambient pressure, CeRhIn$_5$ shows AF ordering at $T<T_{\textrm N}=3.8$
	K and $T_{\textrm N} $ exhibits a moderate variation for 
	$P\siml 1.75$GPa, while 
	the internal field $H_{\textrm{int}}$ is linearly reduced in the region
	$0<P<1.23$GPa, suggesting that $H_{\textrm{int}}$ vanishes 
	at $P\simeq 1.6$GPa.
	At $P \sim 1.6$GPa, the SC state begins to appear,
	even though the AF ordering still exists.
	\cite{Heeger,Mito,SKawasaki,Kitaoka1}
	The detailed NQR measurement shows that 
	at $P=1.75 $GPa, the AF order and the SC order
	coexist on a microscopic level and $1/T_{1}$ in SC state exhibits
	essentially the same $T$-dependence as the normal state
	as is seen in CeCu$_2$Si$_2$.\cite{Mito,SKawasaki,Kitaoka1}
	Under the pressures $P>2.1$GPa, where the AF state disappears,
	the $1/T_1 \propto T^3$ behavior
	is observed as a usual singlet superconductivity with line-nodes.
	\cite{Mito}
	The specific heat of CeRhIn$_5$\cite{Fisher}
	exhibits a feature consistent with the NQR measurement.
	At $P\sim 1.6$GPa, the specific heat shows a broad peak which is 
	expected to be associated with the AF ordering at $T\sim 3$K,
	and shows a shoulder at $T\sim 2$K, which indicates the appearance of the
	gapless superconductivity.
	This shoulder grows to a mean-field like peak at $P\geq 1.9$GPa, 
	which would
	be associated with the anisotropic singlet pairing.
	
\begin{figure}[btp]
	\begin{center}\leavevmode
	\includegraphics[width=0.7\linewidth]{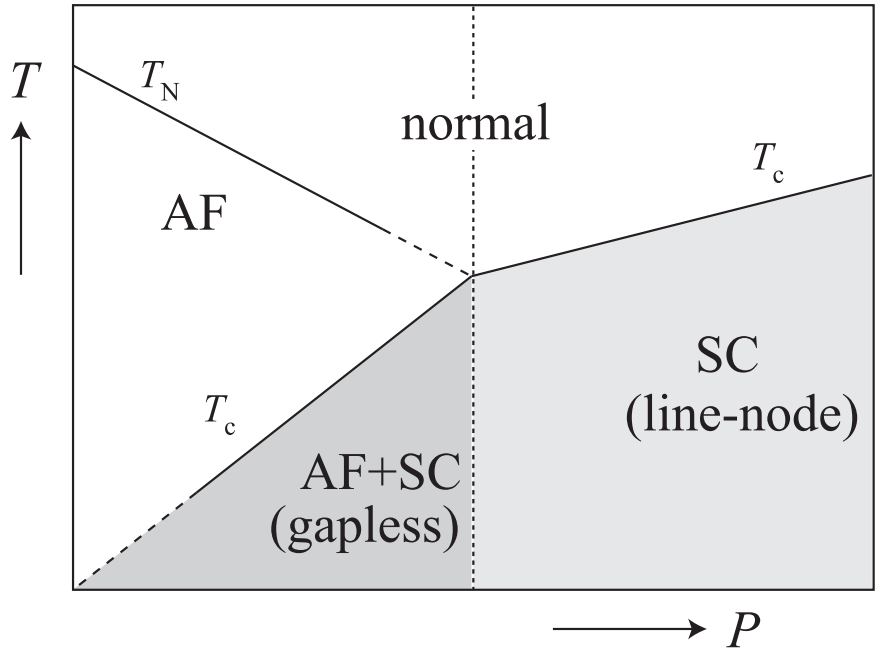}
	\caption{Schematic phase diagram near the phase boundary between
	antiferromagnetic and superconducting states based on the NQR
	experiments on CeCu$_2$Si$_2$ and CeRhIn$_5$.
}\label{phase}\end{center}\end{figure}

	Putting all these together, we can draw a schematic phase diagram
	near the phase boundary between AF and SC states 
	as shown in Fig.\ref{phase}.
	The $P$-$T$ phase diagram can be separated into three parts:
	AF phase, SC phase and a coexisting phase of AF and SC ordering
	(AF+SC phase).
	This coexistence is quite different from that of 
	UPd$_2$Al$_3$,
	in which plural $f$ electrons,
	in $(5f)^3$-configuration of U$^{3+}$ ion,
	shows a dual character of correlated electrons.
	Of three $f$-electrons per U$^{+3}$ ion, 
	two electrons form essentially localized $5f^2$-electron state,
	as in the U$^{4+}$ state with singlet crystalline electric field ground
	state,
	while the remaining one electron forms the heavy Fermi liquid
	due to a larger
	hybridization with conduction electrons.\cite{Sato}
	In Ce-based heavy fermion compounds, on the other hand, 
	there is only one $4f$-electron
	per Ce$^{3+}$ ion, and thus the same $f$-electron exhibits 
	simultaneously itinerant and localized
	dual nature.\cite{Kuramoto}
	The qualitative difference of the manner of the coexistence of AF and SC
	can easily be seen by the property of $1/T_1$ on UPd$_2$Al$_3$\cite{Tou}
	and CeCu$_2$Si$_2$.\cite{Ishida,YKawasaki1,YKawasaki2}

	A crucial aspect of the AF+SC phase is that 
	the gapless superconductivity is realized, 
	although the singlet pairing with line-node due to the 
	antiferromagnetic fluctuation is expected to emerge.
	Although one can soon suspect that the origin of this gapless 
	superconductivity is due to the impurity scattering,
	all the samples above are expected to be very clean. 
	In fact,
	the clear $1/T_1 \propto T^3$ behavior is observed on the same sample
	under the pressures away from the boundary between AF+SC and SC phases.
	Even if the sample is very clean, there is still a possibility of the 
	enhanced impurity scattering associated with quantum critical phenomena.
	\cite{Kotliar,Narikiyo,Maebashi}
	In such a case, however, $\Tc$ should be reduced considerably,
	while Ce$_{0.99}$Cu$_{2.02}$Si$_2$
	at ambient pressure, exhibiting $1/T_1 \propto T$, 
	does not show a clear decrease of $\Tc$.\cite{Ishida,YKawasaki1,SO5}
	Therefore, the AF+SC phase in these Ce-based compounds seem to show 
	the possibility of novel mechanism of ``unconventional" superconductivity.

	In this paper, we present a possible scenario to unravel these 
	mysterious features observed in CeCu$_2$Si$_2$ and CeRhIn$_5$
	near the boundary between AF+SC and SC phases.
	An essential idea here is that 
	a gapless $p$-wave singlet superconductivity with 
	the so-called odd-frequency gap,
	which is odd  both in momentum and in frequency, 
	should be realized very near the quantum critical point (QCP) and/or in
	the AF+SC phase.
	It depends on the specific property of the systems wheather the boundary
	between AF+SC phase and SC phase corresponds to QCP or not.
	However, the present theory can apply to both cases.
	The Pauli principle requires that the spin singlet gap function 
	must be even under simultaneous 
	parity- and time-inversion operations.
	Therefore, there are two types of singlet gap function;
	the first one is even both in frequency and in momentum,
	the other is odd both in frequency and in momentum.
	The latter type of gap was proposed by
	Balatsky and Abrahams (BA) in the context of the research
	of high-$\Tc$ cuprates about a decade ago\cite{Balatsky},
	after the proposal by Berezinskii for the model of superfluid 
	$^3$He.\cite{Berezinskii}
	BA argued that there is no gap in the quasiparticle
	spectrum with odd-frequency gap and these superconductors
	would still exhibit the Meissner effect,
	just as in the case of odd-energy gap
	proposed at very early stage of research 
	as a model of CeCu$_2$Si$_2$ which was claimed to exhibit 
	gapless nature of $1/T_1 $.\cite{Matsuura,Kitaoka2}
	The model phonon interaction discussed by BA was shown to be 
	an impractical one,
	and it was speculated without specifying any explicit models 
	that electron-electron interaction might mediate 
	such odd frequency pairing in general.\cite{Abrahams}

	Here we show that a realistic model interction mediated by
	the critical spin fluctuatioin can give rise to the odd-frequency
	gap superconductivity if the Fermi surface of quasi particles
	satisfies some mild condition.
	In the next section, we describe the model and formalism of 
	solving a linearized gap equation to obtain
	the SC transition temperature $\Tc$.
	In \S 3, we present the numerical results and discuss 
	the condition that $\Tc$ of $p$-wave singlet superconductivity 
	prevails over that of $d$-wave singlet superconductivity.
	In \S 4, we show that the obtained $p$-wave singlet SC state
	is really a gapless state also in real frequency space, 
	i.e., $\Delta (\omega =0)=0$.
	In \S 5, we apply the same method to the region where the AF
	and SC order coexist and show the $p$-wave singlet superconductivity 
	is also realized there.
	Then, we discuss the correspondence between
	results of our theory and experimental results.
	Section 6 is devoted to sammary.

\section{Model and Formulation}
	First we introduce the effective interaction taking into account
	the characteristic features of Ce-based compounds CeCu$_2$Si$_2$
	and CeRhIn$_5$ near the boundary between the AF and SC phase.
	For this purpose, it may be useful to recall the fact that
	the SC transition temperature of CeCu$_2$Si$_2$ (CeRhIn$_5$) exhibits
	maximum at away from the phase boundary $P^* \sim 2.5$GPa
	\cite{Thomas,Holmes} (2.5GPa\cite{}).

	It was shown theoretically, on the standard approximation for an
	extended Anderson lattice model, that
	the maximum of $\Tc$ is raised by the critical 
	valence fluctuations\cite{Onishi}.
	The recent experiment on CeCu$_2$(Si$_{0.9}$Ge$_{0.1}$)$_2$
	has revealed that the SC phase is
	separated into the phase near the phase boundary of AF phase
	and that around $P\sim P^*$\cite{Yuan}.
	So, it is natural to consider that the latter is due to the critical 
	valence fluctuations and the former is due to 
	the critical spin fluctuations.
	This would be also valid for the CeRhIn$_5$ which exhibits 
	the analogous features with the CeCu$_2$Si$_2$.\cite{private}
	Upon this consideration, in order to discuss the mechanism of
	superconductivity near the phase boundary, 
	we introduce the effective interaction 
	$V(\textbf{q}, \textrm{i} \omega_m)$
	as the following phenomenological form mediated by spin fluctuations:
\begin{eqnarray}
	V(\textbf{q}, \textrm{i} \omega_m)=g^2\chi(\textbf{q}, \omega_m)
	\equiv \frac{g^2 N_F}{\eta + A\hat{\textbf{q}}^2 +C|\omega_m |},
	\label{pairint}
\end{eqnarray}
	where $g$ is the coupling constant, $N_F$ the densiyy of states
	at the Fermi level, and $\hat{\textbf{q}}^2\equiv 4+2(\cos q_x +\cos q_y)$
	in two dimensions.
	This type of pairing interaction was adopted by Monthoux and Lonzarich to 
	discuss the strong coupling effect on the superconducticity induced 
	by the critical AF fluctuations.\cite{Monthoux}
	One of the motivations of adopting $\hat{\textbf{q}}^2$ as above is 
	that the ordering vector $\textbf{Q}$ of AF phase in CeRhIn$_5$
	was identified as $\textbf{Q}=(1/2, 1/2, 0.297)$ by the 
	neutron diffraction measurements,\cite{Bao}
	while that in CeCu$_2$(Si$_{0.99}$Ge$_{0.01}$)$_2$ has not been
	identified yet.
	Another one is that calculations become much simpler in two dimensions
	than three dimensions while qualitative physical picture is remained
	unchanged.

\begin{figure}[btp]
	\begin{center}\leavevmode
	\includegraphics[width=0.5\linewidth]{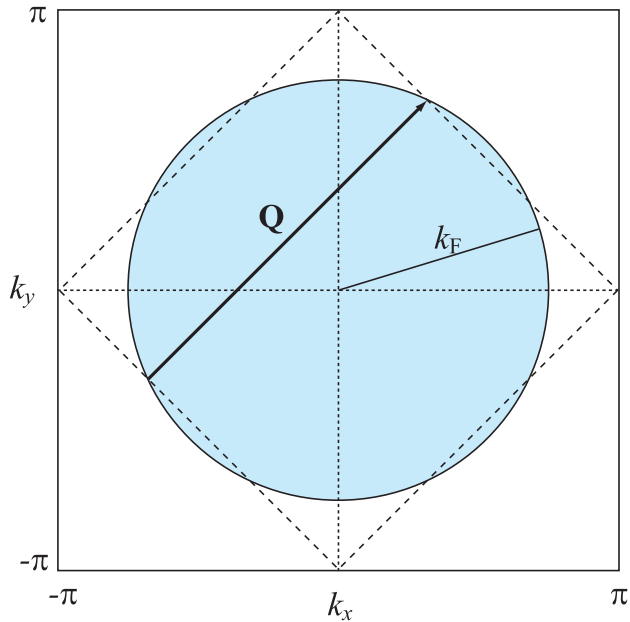}
	\caption{Relation between the Fermi surface ($k_F = 0.75 \pi$)
	and the ordering vector $\textbf{Q}$.
}\label{FS}\end{center}\end{figure}

	The linearized gap equation is given in a weak-coupling approximation
	as follows:
\begin{eqnarray}
	\Delta (\textbf{k}, \textrm{i}\varepsilon_n)
	=-T\sum_{\textbf{k}',\varepsilon_n '}
	\frac{V (\textbf{k}-\textbf{k}',
	\textrm{i}\varepsilon_n -\textrm{i}\varepsilon_n ')}
	{\xi_{\textbf{k}'}^2 + |\varepsilon_n '|^2}
	\Delta (\textbf{k}', \textrm{i}\varepsilon_n '),
\end{eqnarray}
	where $\xi _{\textbf{k}}$ is the dispersion of the quasiparticle,
	and the pairing interaction is given by (\ref{pairint}).
	The pairing interaction can be decomposed as
\begin{eqnarray}
	V (\textbf{k}-\textbf{k}', \textrm{i}\omega_m)=
	\sum_l V_l (\textrm{i}\omega_m )\phi_l^* (\textbf{k})\phi_l (\textbf{k}'),
	\label{irr}
\end{eqnarray}
	where $\phi_l (\textbf{k})$ is an irreducible representation
	of crystal group.
	The coefficient $V_l (\textrm{i}\omega_m)$ in (\ref{irr})
	is given as
\begin{eqnarray}
	V_l (\textrm{i}\omega_m)=\sum_{\textbf{k}, \textbf{k}'}
	\phi_l (\textbf{k}) V (\textbf{k}-\textbf{k}', \textrm{i}\omega_m)
	\phi_l ^*(\textbf{k}') .
\end{eqnarray}
	In order to make our argument as simple and transparent as 
	possible, we concentrate on the frequency dependence of the 
	gap function and assuming the two-dimensional circular Fermi surface
	shown in Fig. \ref{FS}.
	Then, the $p$- and $d$-wave components are retained:
\begin{eqnarray}
	V_l (\textrm{i} \omega_m ) &=& \int \frac{d\textbf{k}}{v_F } 
	\frac{d\textbf{k}'}{v_F }
	\phi_l (\textbf{k} ) V(\textbf{k} -\textbf{k}', \textrm{i}\omega_m )
	\phi_l^* (\textbf{k}), \label{intVl}\\
	\phi_p (\textbf{k}) &\equiv& \Phi_p^{-1}N_F^{-1}
	\delta( \xi _{\textbf{k}}-\mu ) \sin k_{Fx} 
	\,\,(\equiv z_p^{-1}N_F^{-1}\delta( \xi _{\textbf{k}}-\mu )\sin k_{Fy} ),\\
	\phi_d (\textbf{k}) &\equiv& \Phi_d^{-1}N_F^{-1}
	\delta( \xi _{\textbf{k}}-\mu )(\cos k_{Fx} - \cos k_{Fy}),
\end{eqnarray}
	where the wave vector on the Fermi surface are expressed as 
	$\textbf{k}_F = (k_{Fx}, k_{Fy})$,
	and $\Phi_l$ is the normalization factor.
	
	The linearized gap equation for each partial-wave component
	 can be written in the form:
\begin{eqnarray}
	\lambda (T)\Delta_l (\textrm{i}\varepsilon_n)
	=-T\sum_{\textbf{k}',\varepsilon_n '}
	\frac{V_l ( \textrm{i}\varepsilon_n -\textrm{i}\varepsilon_n ')}
	{\xi_{\textbf{k}'}^2 + |\varepsilon_n '|^2}
	\Delta_l (\textrm{i}\varepsilon_n ').
	\label{gapeq}
\end{eqnarray}
	Assuming $\xi_{\textbf{k}}=v_F (|\textbf{k}|-k_F )$,
	one can carry out the integration
	of eq. (\ref{gapeq}) with respect to $\textbf{k}'$
	as follows:
\begin{eqnarray}
	\sum_{\textbf{k}}
	\frac{1}{\xi_{\textbf{k}}^2 + |\varepsilon_n |^2}
	&=&\frac{1}{(2\pi )^2}\int \textrm{d}\textbf{k}
	\frac{1}{v_F^2 (|\textbf{k}|-k_F)^2 +|\varepsilon_n |^2} \nonumber \\
	&\simeq &\frac{1}{2\pi }\int_{k_F -k_0}^{k_F + k_0} \textrm{d}k_F
	\frac{k}{v_F^2 (k-k_F)^2 +|\varepsilon_n |^2} \nonumber \\
	&=&\frac{k_F}{\pi v_F |\varepsilon_n |}
	\tan ^{-1} \frac{v_F k_0}{|\varepsilon_n |}.
\end{eqnarray}

	As mentioned above, 
	the gap function must satisfy the following symmetry 
	for the spin-singlet pairing
\begin{eqnarray}
	\Delta_d (\textbf{k}, \textrm{i}\varepsilon_n)
	&=&\Delta_d (-\textbf{k}, \textrm{i}\varepsilon_n)
	=\Delta_d (\textbf{k}, -\textrm{i}\varepsilon_n)
	=\Delta_d (-\textbf{k}, -\textrm{i}\varepsilon_n), \label{even}\\
	\Delta_p (\textbf{k}, \textrm{i}\varepsilon_n)
	&=&-\Delta_p (-\textbf{k}, \textrm{i}\varepsilon_n)
	=-\Delta_p (\textbf{k}, -\textrm{i}\varepsilon_n)
	=\Delta_p (-\textbf{k}, -\textrm{i}\varepsilon_n).	\label{odd}
\end{eqnarray}
	Here we measure energies in unit of the Fermi energy $\varepsilon_F $.
	$g^2 N_F $ is set to be equal to $\varepsilon_F$ throughout the paper.
	The Fermi surface (FS) is assumed to be two-dimensional circular
	with $k_F = 0.75 \pi$ as is shown in Fig. \ref{FS}.
	This simplified model may give clear understanding of a condition
	for the realization
	of the odd-frequency gap, although it does not corresponds to
	the practical Fermi surface of the heavy fermion compounds.

\begin{figure}[btp]
	\begin{center}\leavevmode
	\includegraphics[width=0.7\linewidth]{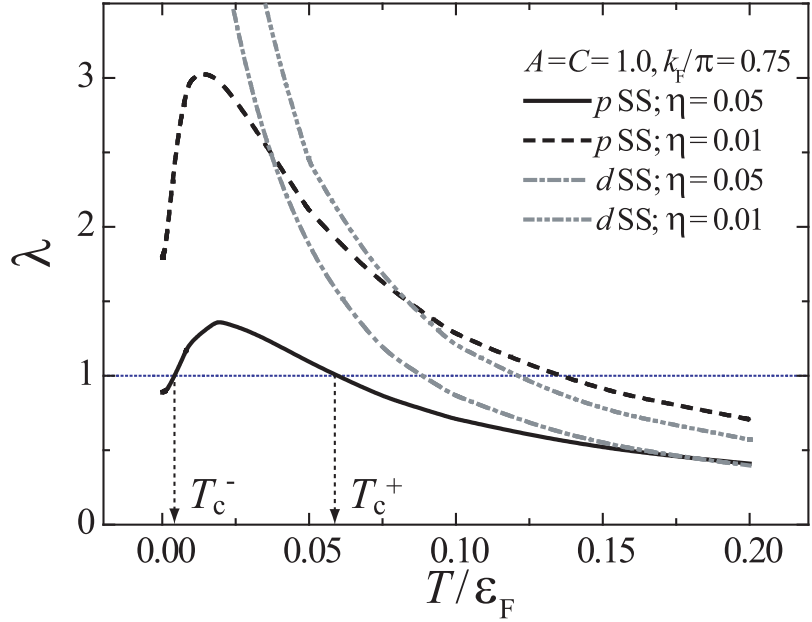}
	\caption{Temperature dependence of the eigen value $\lambda $
	for $p$-wave singlet ($p$-S) and $d$-wave singlet ($d$-S).
}\label{lambda}\end{center}\end{figure}

	The transition temperature $\Tc$ is determined by condition
	$\lambda (\Tc )=1$.
	We have solved the eigenvalue problem (\ref{gapeq}) numerically 
	by retainning 512 Matsubara frequency $\varepsilon _n$
	up to $|\varepsilon _n |\le 131583\pi T$.
	The temperature dependences of the eigen value $\lambda (T)$'s
	are shown in Fig. \ref{lambda}.
	For the $d$-wave pairing, $\lambda (T)$ increases monotonously 
	(logarithmically) as the temperature decreases as usual.
	For the $p$-wave odd-frequency pairing, on the other hand, $\lambda (T)$
	exhibits a broad peak as shown in Fig. \ref{lambda}.
	Since this feature can be seen even in the case of the much larger
	cut-off frequency, it is not due to
	the cut-off frequency, but due to an intrinsic nature of the odd-pairing
	in frequency.
	Consequently, it is possible that there exist two temperatures $\Tc ^{\pm}$
	satisfing $\lambda (\Tc ^{\pm})=1$,
	so that the $p$-wave pairing is realized only in the temperature
	region $\Tc ^- < T<\Tc ^+$ (see the case 
	$\eta = 0.05$ in Fig. \ref{lambda}).
	But in the case $\eta < 0.04$, $\lambda$ does not fall below unity
	even for $T\to 0$, so that the $p$-wave pairing is realized
	in $0\le T<\Tc ^+$.

	We recall that BA reached the anologous results
	for the odd-frequency pairing
	that the normal phase reappears below the lower-$\Tc$.
	In the present model, we cannot show the existence of $\Tc^{\pm}$,
	since 
	the transition temperature for $d$-wave singlet pairing $\Tc^d$ is
	higher than $\Tc^+$ for $\eta >0.04$.
	The detailed analysis about the possibility of 
	this multiple transition will be
	given in a subsequent paper.\cite{Fuseya}

\section{Competition between $p$-wave and $d$-wave pair states}
\begin{figure}[btp]
\begin{center}\leavevmode
	\includegraphics[width=0.7\linewidth]{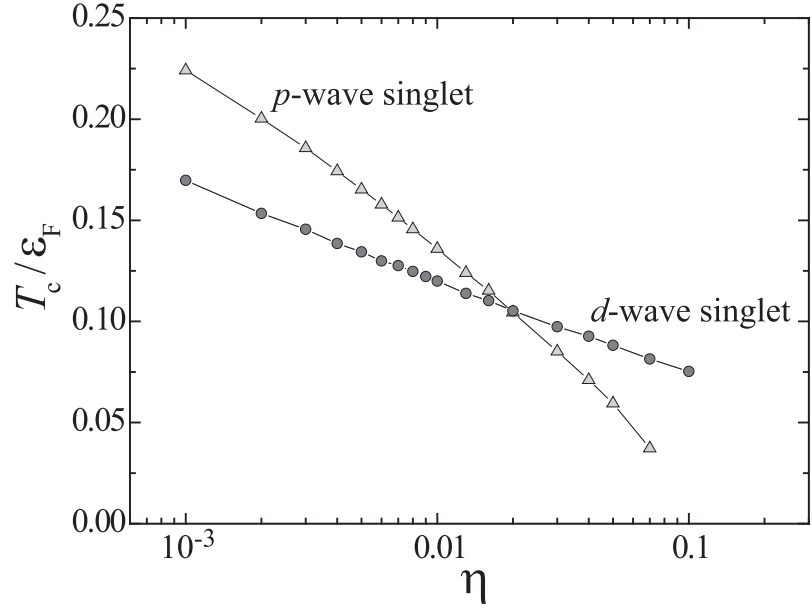}
	\caption{Transition temperature for $p$($d$)-wave singlet pairing, 
	$\Tc^{p (d)}$, as a function of $\eta$.
}\label{Tceta}\end{center}\end{figure}

	As is shown in Fig. \ref{Tceta}, the transition temperature
	for $p$-wave singlet $\Tc^p$ exceeds
	$\Tc^d$ when $\eta \siml 0.02$.
	That is to say, in the region away from the QCP,
	$d$-wave singlet pairing is realized and $p$-wave singlet pairing
	is formed very close to the QCP.
\begin{figure}[btp]
\begin{center}\leavevmode
	\includegraphics[width=0.7\linewidth]{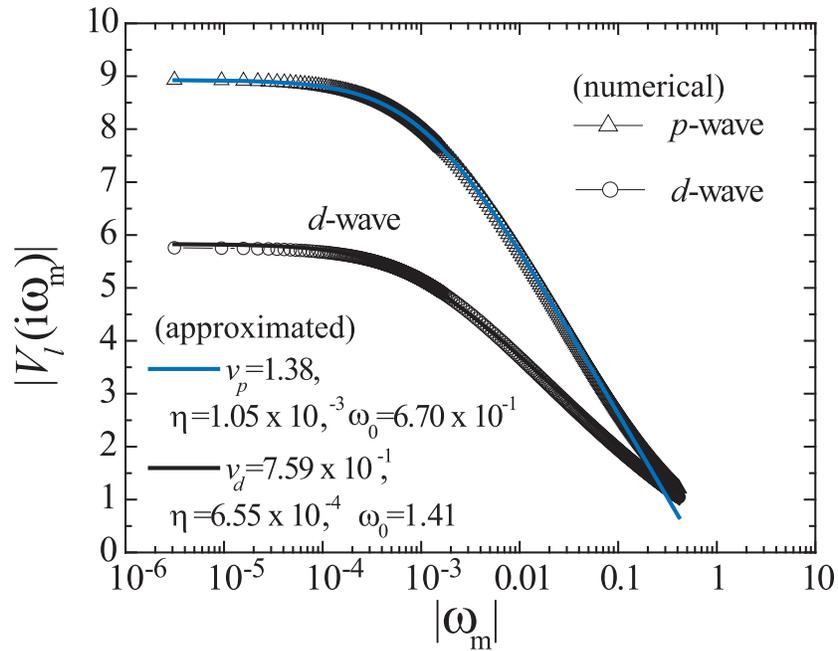}
	\caption{Frequency dependence of $V_l (\omega _m )$
}\label{Vl}\end{center}\end{figure}

	The reason why such $p$-wave singlet pairing arises
	near the AF-QCP can be understood 
	as follows.
	The frequency dependence of $V_l (\textrm{i}\omega_m)$, which
	is calculated numerically by eq. (\ref{intVl}), are shown in 
	Fig. \ref{Vl}.
	As is also shown in Fig. \ref{Vl}, 
	it should be reasonable to approximate 
	the frequency dependence of $V_l (\textrm{i} \omega_m)$ as
\begin{eqnarray}
	V_l (\textrm{i} \omega_m) \simeq v_l \ln \frac{\omega_0}{\tilde{\eta} 
	+ |\omega_m |}.
	\label{appVl}
\end{eqnarray}
	Considering the symmetry property of the gap function, eqs. (\ref{even}) 
	and (\ref{odd}),
	we can express each $V_l$ in the form:
\begin{eqnarray}
	V_p (\textrm{i}\varepsilon_n -\textrm{i}\varepsilon_n')&=&\frac{v_p}{2}
	\Biggl( \ln \frac{1}{\tilde{\eta} +|\varepsilon_n -\varepsilon_n'|}
	-\ln \frac{1}{\tilde{\eta} +|\varepsilon_n +\varepsilon_n'|}
	\Biggr), \\
	V_d (\textrm{i}\varepsilon_n -\textrm{i}\varepsilon_n')&=&\frac{v_d}{2}
	\Biggl( \ln \frac{1}{\tilde{\eta} +|\varepsilon_n -\varepsilon_n'|}
	+\ln \frac{1}{\tilde{\eta} +|\varepsilon_n +\varepsilon_n'|}
	\Biggr).
\end{eqnarray}

	Let us consider the most singular case, $\varepsilon_n = \varepsilon_n'=\pi T$.
	Of course this example does not reproduce the present situation exactly, 
	but gives a simple picture of the competition of the pairing states.
	Each $V_l$ is given by
\begin{eqnarray}
	V_p (0) &=& \frac{v_p}{2}\ln \frac{\tilde{\eta}+2\pi T}{\tilde{\eta}}, \\
	V_d (0) &=& \frac{v_d}{2}\ln \frac{1}{\tilde{\eta}(\tilde{\eta}+2\pi T)}.
\end{eqnarray}
	When the system locates away from the QCP, i.e., $\tilde{\eta}\gg 2\pi T$,
\begin{eqnarray}
	V_p (0) &\simeq& \frac{v_p}{2}\frac{2\pi T}{\tilde{\eta}}, \\
	V_d (0) &\simeq& \frac{v_d}{2}\Biggl( -2\ln \tilde{\eta}
	-\frac{2\pi T}{\tilde{\eta}}\Biggr).
\end{eqnarray}
	In this case, $V_p $ is almost negligeble compared to $V_d$,
	leading to $\Tc^d \gg \Tc^p$.

	In contrast, when the system locates very close to the QCP, i.e.,
	$\tilde{\eta}\ll 2\pi T $,
\begin{eqnarray}
	V_p (0) &\simeq& \frac{v_p}{2}\Biggl( 
	\frac{\tilde{\eta}}{2\pi T}-\ln \frac{\tilde{\eta}}{2\pi T}
	\Biggr) \sim  -\frac{v_p}{2}\ln \frac{\tilde{\eta}}{2\pi T}, \\
	V_d (0) &\simeq& \frac{v_d}{2}\Biggl( -2\ln (2\pi T)-\frac{\tilde{\eta}}{2\pi T}
	-\ln \frac{\tilde{\eta}}{2\pi T}
	\Biggr) \sim -\frac{v_d}{2}\ln \frac{\tilde{\eta}}{2\pi T}. 
\end{eqnarray}
	This time $V_p$ is comparable to $V_d$, and in the case of
	$|v_p| > |v_d|$, $\Tc^p$ becomes higher than $\Tc^d$.
	
\begin{figure}[btp]
\begin{center}\leavevmode
	\includegraphics[width=0.7\linewidth]{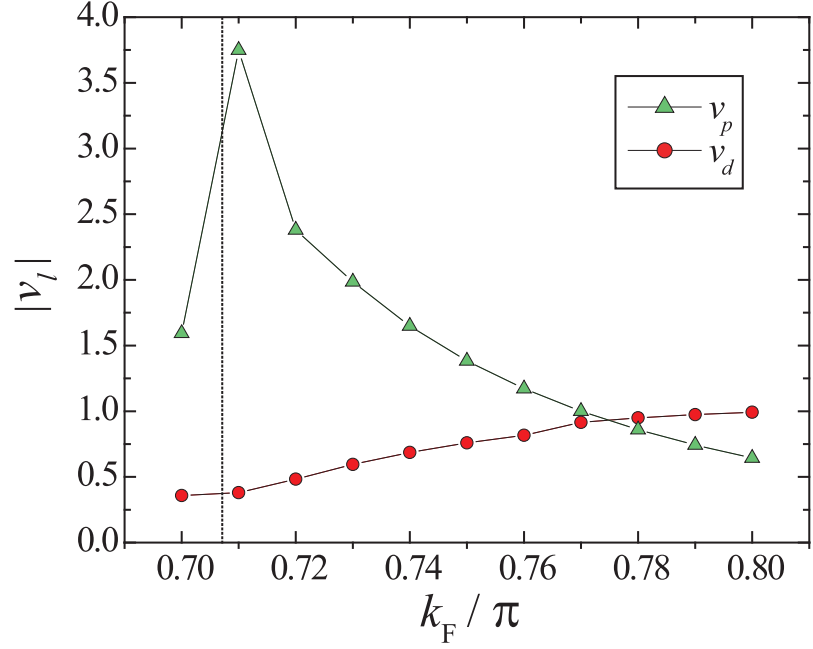}
	\caption{$k_F$-dependence of $v_l $.
	The vertical dotted-line indicates $k_F =\pi/\sqrt{2}$ where
	$|\textbf{Q}|$ is equivalent to the diameter of the Fermi surface.
}\label{Vl0}\end{center}\end{figure}
	
\begin{figure}[btp]
\begin{center}\leavevmode
	\includegraphics[width=1.0\linewidth]{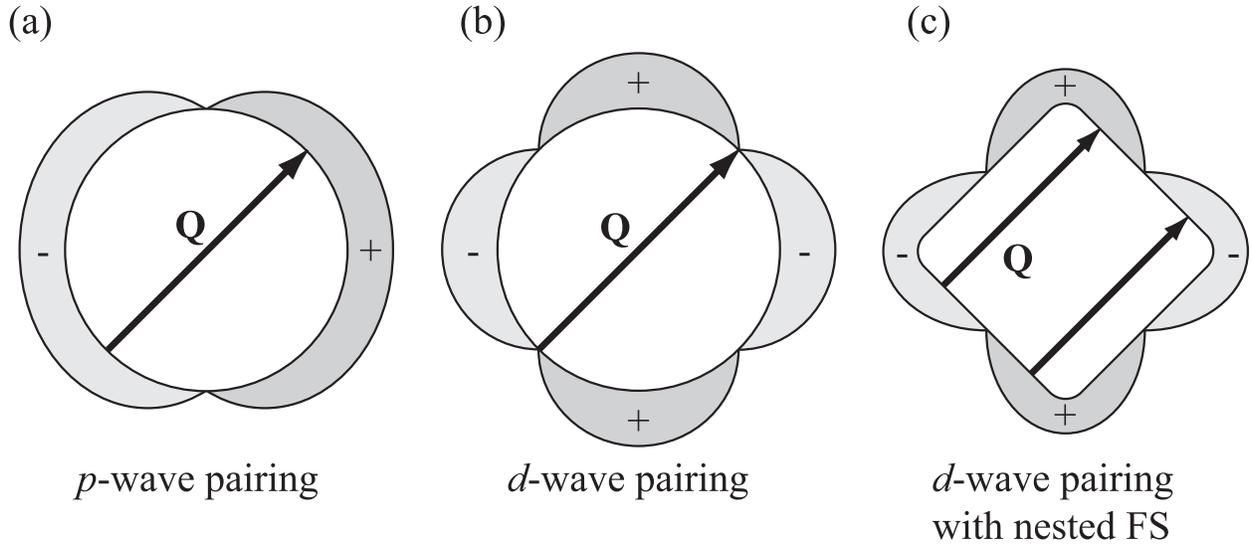}
	\caption{Illustration of the relation between the AF ordering vector
	$\textbf{Q}$ and the position of the nodes.
	(a) $p$-wave pairing case: 
	the most singlular pair scattering with $textbf{Q}$
	is not suppressed by the existence of the nodes.
	(b) $d$-wave pairing case: the scattering with $\textbf{Q}$ is suppressed
	by the existence of the nodes.
	(c) $d$-wave pairing with FS nested with $\textbf{Q}$ case:
	the suppression of pair scattering becomes small.
	
}\label{pdH}\end{center}\end{figure}

	As shown in Fig. \ref{Vl0}, $|v_l|$ considerably depends on $k_F$.
	In other words,
	it is determined by the relation between the Fermi surface
	and AF ordering vector $\textbf{Q}$.
	In the present case,
	the most singular pair scattering with $\textbf{q}=\textbf{Q}=
	(\pm \pi, \pm \pi)$
	uses the points near $(\pm k_F /\sqrt{2}, \pm k_F /\sqrt{2})$, where
	the nodes of the $d$-wave pairing exists while that of the 
	$p$-wave pairing does not. (See Fig. \ref{pdH} (a), (b))
	Therefore the effects of the singular pair scattering are 
	suppressed in the case of $d$-wave pairing.
	This causes the relation $|v_p |>|v_d |$ in the present case.

	Considering the above discussion, we can extend the condition
	of the emergence of the $p$-wave singlet pairing to the general case
	as following.
	Firstly, the pair scattering interaction must have a sharp peak in
	frequency whose width is smaller than the temperature (in the present
	case, $\eta \ll T$).
	Secondly, the most singular pair scattering with vector $\textbf{Q}$
	has to be canceled out by the nodes of the $d$-wave pairing.
	In the case of Hubbard model at the half-filling, for example,
	the Fermi surface is nested with $\textbf{Q}=(\pm \pi, \pm \pi)$
	and the suppression of pair scattering becomes small,
	namely, $|v_p | \sim |v_d |$ (see Fig \ref{pdH} (c)).
	So, it would be difficult for the $p$-wave singlet pairing 
	to prevail over the $d$-wave singlet pairing on the Hubbard model,
	even near the AF-QCP.
	How about the cases of CeCu$_2$Si$_2$ and CeRhIn$_5$?
	The $1/T_1$ of both compounds rapidly increase approaching
	the AF+SC phase from SC phase.\cite{YKawasaki1,SKawasaki}
	This indicates that the system is located very near the QCP,
	which would satisfy the first condition mentioned above.
	In CeRhIn$_5$, the most singular pair scattering vector is 
	$\textbf{Q}=(\pi, \pi, 0.297\pi)$\cite{Bao}.
	Its Fermi surface does not nest in the direction $\textbf{Q}$
	and is satisfied the relation
	$\textbf{k}_F - \textbf{k}'_F =\textbf{Q}$ 
	in the direction $(\pm 1, \pm 1, 0)$.
	\cite{Settai}
	The feature that $\textbf{Q}$ is independent from the nesting vector
	can be understood from the dual nature of electron.
	According to the itinerant-localized duality model\cite{Kuramoto},
	the spin susceptibility $\chi (q, \textrm{i}\omega_m)$ is
	expressed in terms of the polarization function 
	$\Pi (q, \textrm{i}\omega_m)$  of quasiparticles
	and the exchange interaction $J(q, \textrm{i}\omega_m)$ between
	localized component of electrons as follows:
\begin{eqnarray}
	\chi (q, \textrm{i}\omega_m)^{-1}
	=\chi_0 (\textrm{i}\omega_m)^{-1} -2\lambda^2 \Pi (q, \textrm{i}\omega_m)
	-J(q, \textrm{i}\omega_m),
\end{eqnarray}
	where $\chi_0$ is the local susceptibility expressing an effect of 
	Kondo-like correlation, and $\lambda$ is the renormalized 
	spin-fermion coupling.
	Even if the Fermi surface is not nested, i.e., $\Pi $ is not enhanced,
	the AF ordering can be triggerd by $J(\textbf{Q})$.
	Therefore, CeRhIn$_5$ would satisfy both conditions.
	In the case of CeCu$_2$Si$_2$, a situation is much more complicated 
	($\textbf{Q}\sim (0.27\pi, 0.27\pi, 0.52\pi)$\cite{Knebel}),
	but its Fermi surface seems to satisfy the condition.\cite{Harima}

\section{Gaplessness of odd-frequency pairing}

\begin{figure}[btp]
\begin{center}\leavevmode
	\includegraphics[width=0.7\linewidth]{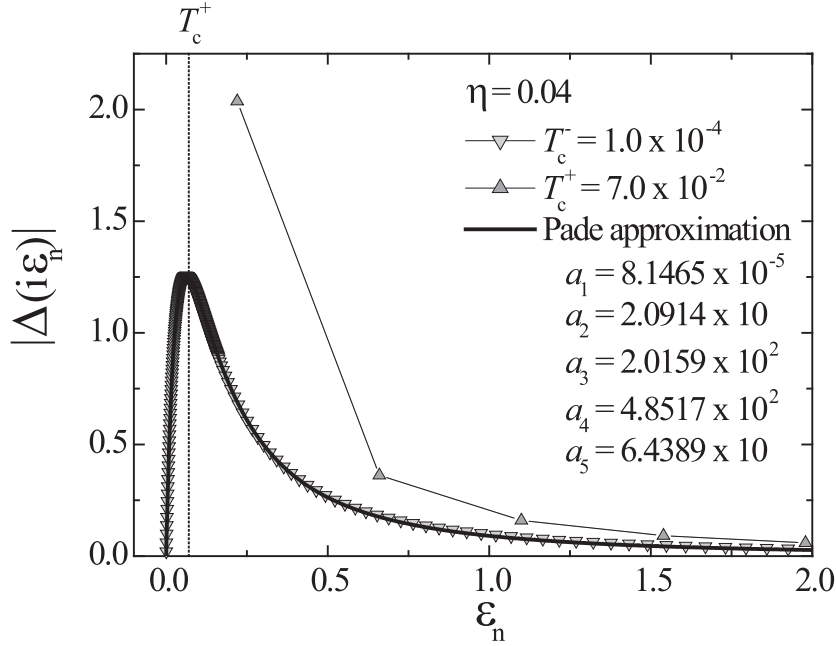}
	\caption{Frequency dependence of the gap function of the $p$-wave singlet
	pairing at $\Tc^+ /\varepsilon _F = 7.0 \times 10^{-2}$
	and $\Tc^- /\varepsilon _F = 1.0 \times 10^{-4}$for $\eta=0.04$.
	The solid line indicates the results of Pad\'e approximation.
}\label{gap}
\end{center}
\end{figure}

	The frequency dependence of the gap function of the $p$-wave singlet
	pairing at $\Tc^+  = 7.0 \times 10^{-2}$ and $\Tc^- 
	= 1.0 \times 10^{-4}$ for $\eta=0.04$
	are shown in Fig. \ref{gap}.
	For $T=\Tc^+$, at first sight, the gap seems to be divergent
	when $\varepsilon _n$ approaches zero.
	The gap at $T=\Tc^-$, on the other hand, has a peak at $\varepsilon_n^* =
	6.19 \times 10^{-2}$, which is comparable to $\Tc^+ $,
	and vanishes for $\varepsilon \to 0$, i.e., 
	it exhibits a gapless-like structure.
	This feature seems robust against variations of 
	$\eta , k_F$ or $g$.
	Therefore, we can suspect that the peak and the gapless-like structure is
	only hidden for $T=\Tc^+$.

	At the present stage, however, we cannot declare wheather the present
	$p$-wave singlet gap function is completely gapless or not.
	So, we have to perform the analytic continuation of the gap function.
	In order to do that, first we apply the Pad\'e approximation
	to the calculated gap function.
	Since the gap function is expected to be proportional to 
	$|\varepsilon_n |^{-2}$
	for large $\varepsilon_n $,
	we take the form:
\begin{eqnarray}
	\Delta_{\textrm{P}} (\textrm{i}\varepsilon_n )
	=\frac{a_5 (\varepsilon_n + a_1 )}{1+ a_2 \varepsilon_n 
	+a_3 \varepsilon_n^2 + a_4 \varepsilon_n^3},
\end{eqnarray}
	for $\varepsilon _n >0$.
	We apply the least squares method to the calculated gap function 
	$\Delta (\textrm{i}\varepsilon_n )$by 
	$\Delta_{\textrm{P}}(\textrm{i}\varepsilon_n )$ and obtain
	$a_1 = -1.2455 \times 10^{-5}, a_2 =2.2966\times 10, 
	a_3 = 1.6980 \times 10^2 , a_4 =7.7980 \times 10^2 ,
	a_5 = 6.6046 \times 10$.
	Using these values of $a_{1\sim 5}$, we display
	$\Delta_{\textrm{P}}(\textrm{i}\varepsilon_n )$ in Fig. \ref{gap},
	in which $\Delta_{\textrm{P}}(\textrm{i}\varepsilon_n )$ 
	shows a good agreement with $\Delta (\textrm{i}\varepsilon_n )$.
	The analytic continuation of this form becomes
\begin{eqnarray}
	\textrm{Re} \Delta_{\textrm{P}}(\varepsilon )&=&
	\frac{a_1 a_5 (1-a_3 \varepsilon^2 )+a_5 \varepsilon^2 
	(a_2-a_4 \varepsilon^2 )}{(1-a_3 \varepsilon^2 )^2
	+\varepsilon^2 (a_2 -a_4 \varepsilon^2)}, \label{PadeRe}\\
	\textrm{Im} \Delta_{\textrm{P}}(\varepsilon )&=&
	\frac{a_5 \varepsilon (a_3 \varepsilon^2 -1)+a_1 a_5 \varepsilon
	(a_2-a_4 \varepsilon^2 )}{(1-a_3 \varepsilon^2 )^2
	+\varepsilon^2 (a_2 -a_4 \varepsilon^2)}. \label{PadeIm}
\end{eqnarray}
	From eqs. (\ref{PadeRe}) and (\ref{PadeIm}), the gap function
	at $\varepsilon = 0$ is reduced to
	$\textrm{Re} \Delta_{\textrm{P}}(0)=-8.2260 \times 10^{-4}, \,\,
	\textrm{Im} \Delta_{\textrm{P}}(0)=0$.
	Therefore, we can conclude the present $p$-wave singlet superconductivity
	is a technically gapless one.
	We can determine the maximum of the gap, $\Delta_0 (T)$ below $\Tc$,
	by assuming that the gap has the same frequency dependence
	as that of the eigenfunction at $T \ll \Tc$, where the eigenvalue $\lambda$
	exceeds unity further.
	Namely,
	$\Delta (\textrm{i}\varepsilon_n ; T) \equiv \Delta_0 (T)
	\Delta_P (\textrm{i}\varepsilon_n ; T\ll\Tc)$.
	At T=0, we obtained $\Delta_0 (0) / \Tc = 0.42$ (by setting parameters 
	as $a_1 = -1.5256 \times 10^{-5}, a_2 = 4.6835 \times 10, 
	a_3 = 6.4780 \times 10^2, a_4 = 2.8209 \times 10^3, 
	a_5 = 1.1492 \times 10^2$, which corresponds to $\eta = 0.01$ case), 
	which is much smaller
	than the BCS relation $\Delta_0 (0) / \Tc = 1.765$.

	We can also discuss the gapless nature in more general form as follows.
	The linearized gap equation for the odd-frequency gap
	can be written as 
\begin{eqnarray}
	\Delta_p (\textrm{i}\varepsilon_n ) &=&-N_F \int_{\infty}^{\infty}
	\textrm{d}\xi T\sum_{\varepsilon_n'}
	\frac{V(\textrm{i}\varepsilon_n -\textrm{i}\varepsilon_n ')}
	{|\varepsilon_n'|^2 +\xi ^2}
	\frac{1}{2}\bigl[ \Delta_p (\textrm{i}\varepsilon_n' )
	-\Delta_p (-\textrm{i}\varepsilon_n' ) \bigr] \\
	&=&-N_F \textrm{P}\int_{\infty}^{\infty}
	\frac{\textrm{d}x}{8}
	\coth \frac{x}{2T}\frac{1}{x+\textrm{i}\varepsilon_n}
	\bigl[ V(-x-\textrm{i}\delta )-V(-x+\textrm{i}\delta ) \bigr]
	\times 
	\bigl[ \Delta (\textrm{i}\varepsilon_n +x)-
	\Delta (-\textrm{i}\varepsilon_n -x)
	\bigr] \nonumber \\
	& &-N_F \textrm{P}\int_{\infty}^{\infty}
	\frac{\textrm{d}x}{8}
	\tanh \frac{x}{2T}\frac{1}{x+\textrm{i}\delta}
	\bigl[ V(\textrm{i}\varepsilon_n -x)-V(\textrm{i}\varepsilon_n +x) \bigr]
	\times 
	\bigl[ \Delta (x+\textrm{i}\delta)-\Delta (-x-\textrm{i}\delta ) 
	\bigr].
	\label{gapeqM}
\end{eqnarray}
	The analytic continuation, $\textrm{i} \varepsilon_n \to 
	\varepsilon + \textrm{i}\delta$
	in eq. (\ref{gapeqM}), results in the following gap equation:
\begin{eqnarray}
	\Delta (\varepsilon +\textrm{i}\delta )
	&=&-N_F \textrm{P}\int_{\infty}^{\infty}
	\frac{\textrm{d}x}{8}
	\Bigl\{
	\coth \frac{x-\varepsilon }{2T}
	\bigl[ V(\varepsilon -x -\textrm{i}\delta )-
	V(\varepsilon -x +\textrm{i}\delta )
	\bigr] \nonumber \\
	& &\qquad \qquad \qquad \qquad +\tanh \frac{x}{2T}
	\bigl[ V(\varepsilon -x +\textrm{i}\delta )
	-V(\varepsilon +x +\textrm{i}\delta )
	\bigr] \Bigr\} \nonumber \\
	& &\qquad \qquad \qquad
	\times \frac{\Delta (x+\textrm{i}\delta )-\Delta (-x-\textrm{i}\delta )}
	{x+\textrm{i}\delta}.
	\label{anagap}
\end{eqnarray}
	
	The analytic continuation of the effective pairing interaction
	$V_l (\textrm{i}\omega_m )$, (\ref{appVl}), becomes
\begin{eqnarray}
	V(\omega +\textrm{i}\delta )=v_l 
	\bigl( \ln \frac{\omega_0}{\sqrt{\omega ^2 +\tilde{\eta}^2}}
	+\textrm{i} \tan ^{-1} \frac{\omega }{\tilde{\eta}}
	\bigr)
	\label{anaVl}
\end{eqnarray}
	Substituting eq. (\ref{anaVl}) into eq. (\ref{anagap}),
	we obtain 
\begin{eqnarray}
	\Delta (\varepsilon +\textrm{i}\delta ) &=&
	-v_l N_F \textrm{P} \int_{-\infty}^{\infty}
	\frac{\textrm{d}x}{8}
	\Bigl\{
	\tanh \frac{x}{2T}\ln\sqrt{\frac{(\varepsilon +x )^2 +\tilde{\eta}^2}
	{(\varepsilon -x )^2 +\tilde{\eta}^2}} \nonumber\\
	& &+\textrm{i} \bigl[ 
	2\coth \frac{x-\varepsilon }{2T}
	\tan ^{-1} \frac{\varepsilon -x}{\tilde{\eta}}
	+\tanh \frac{x}{2T} 
	\bigl(
	\tan ^{-1} \frac{\varepsilon -x}{\tilde{\eta}}
	-\tan ^{-1} \frac{\varepsilon +x}{\tilde{\eta}}
	\bigr)
	\bigr] \Bigr\} \nonumber \\
	& &\qquad \qquad \qquad \qquad \qquad 
	\times \frac{\Delta (x+\textrm{i}\delta )-\Delta (-x-\textrm{i}\delta )}
	{x+\textrm{i}\delta}.
	\label{anagapeq}
\end{eqnarray}

	Let us express the real (imaginary) part of the gap as 
	$\Delta '(\varepsilon )$ ($\Delta '' (\varepsilon )$),
	i.e., $\Delta (\varepsilon +\textrm{i}\delta ) \equiv 
	\Delta '(\varepsilon ) +\textrm{i}\Delta ''(\varepsilon )$.
	Then, each part of (\ref{anagapeq}) are reduced to
\begin{eqnarray}
	\Delta '(\varepsilon)
	&=&-v_l N_F \textrm{P}\int_{-\infty}^{\infty}
	\frac{\textrm{d}x}{8}
	\Biggl\{
	\tanh \frac{x}{2T} \ln\sqrt{\frac{(\varepsilon +x )^2 +\tilde{\eta}^2}
	{(\varepsilon -x )^2 +\tilde{\eta}^2}}
	\frac{\Delta '(x)-\Delta '(-x)}{x} \nonumber \\
	& &-\Bigl[
	2\coth \frac{\varepsilon -x}{2T} \tan ^{-1}
	\frac{\varepsilon -x}{\tilde{\eta}}
	+\tanh \frac{x}{2T}
	\bigl(
	\tan ^{-1} \frac{\varepsilon -x}{\tilde{\eta}}
	-\tan ^{-1} \frac{\varepsilon +x}{\tilde{\eta}}
	\bigr)
	\Bigr]
	\frac{\Delta ''(x)+\Delta ''(-x)}{x}
	\Biggr\} \nonumber \\
	\\
	\Delta '' (\varepsilon)
	&=&-v_l N_F \textrm{P}\int_{-\infty}^{\infty}
	\frac{\textrm{d}x}{8}
	\Biggl\{
	\tanh \frac{x}{2T} \ln\sqrt{\frac{(\varepsilon +x )^2 +\tilde{\eta}^2}
	{(\varepsilon -x )^2 +\tilde{\eta}^2}}
	\frac{\Delta ''(x)+\Delta ''(-x)}{x} \nonumber \\
	& &-\Bigl[
	2\coth \frac{\varepsilon -x}{2T} \tan ^{-1}
	\frac{\varepsilon -x}{\tilde{\eta}}
	+\tanh \frac{x}{2T}
	\bigl(
	\tan ^{-1} \frac{\varepsilon -x}{\tilde{\eta}}
	-\tan ^{-1} \frac{\varepsilon +x}{\tilde{\eta}}
	\bigr)
	\Bigr] \nonumber \\
	& & \times \biggl[ \frac{\Delta '(x)-\Delta '(-x)}{x}
	+\bigl\{ \Delta ''(x) +\Delta ''(-x) \bigr\} \pi \delta (x)
	\biggr]\Biggr\}.
\end{eqnarray}
	Note here that we can show easily 
	$\Delta '(\varepsilon )=-\Delta '(-\varepsilon)$,
	$\Delta (\varepsilon )=\Delta ''(-\varepsilon)$.
	At $\varepsilon =0$, 
\begin{eqnarray}
	\Delta '(0) &=&0, \label{Delta'0}\\
	\Delta ''(0) &=& -v_l N_F \textrm{P}\int_{-\infty}^{\infty}
	\frac{\textrm{d}x}{2}
	\biggl( \coth \frac{x}{2T}-\tanh \frac{x}{2T}
	\biggr)
	\tan ^{-1}\frac{x}{\tilde{\eta}} \cdot
	\frac{\Delta '(x)}{x}
	-v_l N_F \pi \frac{T}{\tilde{\eta}}\Delta ''(0).
	\label{Delta''0}
\end{eqnarray}
	From eq. (\ref{Delta''0}),
\begin{eqnarray}
	\biggl(1+v_l N_F \pi \frac{T}{\tilde{\eta}}
	\biggr) \Delta ''(0) = -v_l N_F \textrm{P}\int_{-\infty}^{\infty}
	\frac{\textrm{d}x}{2}
	\biggl( \coth \frac{x}{2T}-\tanh \frac{x}{2T}
	\biggr)
	\tan ^{-1}\frac{x}{\tilde{\eta}} \cdot
	\frac{\Delta '(x)}{x}.
	\label{Delta''02}
\end{eqnarray}
	The right hand side of eq. (\ref{Delta''02}) is constant,
	and $1+v_l N_F \pi T/\tilde{\eta} \to -\infty $
	when $\tilde{\eta}\to 0$.
	Therefore, in order to satisfy eq. (\ref{Delta''02}),
\begin{eqnarray}
	\Delta ''(0) \to 0 \hspace{1cm} \textrm{for} 
	\hspace{1cm} \tilde{\eta}\to 0
	\label{Delta''03}
\end{eqnarray}
	is needed.
	From eqs. (\ref{Delta'0}) and (\ref{Delta''03}),
	we reach the fact that the present $p$-wave
	singlet gap function leads to an essentially gapless superconductivity
	at the QCP.

	Considering the low-frequency structure of the gap function,
	we assume that the gap function can be approximate in the form:
\begin{eqnarray}
	\Delta (\textbf{k}, \textrm{i}\omega_m )= \frac{\Delta_0}{\Tc} 
	\textrm{i}\omega_m \phi_p (\textbf{k}).
\end{eqnarray}
	Then we find from the poles of Green's function that
\begin{eqnarray}
	E_{\textbf{k}}=\frac{\xi}{\sqrt{1+(\Delta_0 /\Tc)^2 \phi_p^2 
	(\textbf{k})}}.
\end{eqnarray}
	Therefore, the quasiparticle spectra with such gap function is 
	gapless, i.e., there is no difference in excitations between 
	the nomal states and the present SC states,
	except for the effective mass enhancement:
\begin{eqnarray}
	m_{\textbf{k}}^* = m\sqrt{1+(\Delta_0 /\Tc)^2 \phi_p^2 (\textbf{k})}.
\end{eqnarray}
	This mass enhancement is, however, very small
	due to the smallness of $\Delta_0 /\Tc \sim 0.42$ as mentioned above, 
	so that the specific heat or the NMR relaxation rate does not
	show any significant change at $\Tc$.
	This would correspond to the $1/T_1$ behavior 
	observed in Ce$_{0.99}$Cu$_{2.02}$Si$_2$
	\cite{Ishida,YKawasaki1,SO5}
	which shows almost the same behavior as the normal Fermi liquid state.
	If the $1/T_1 \propto T$ behavior is due to the impurity scattering,
	$1/T_1 $ should shows a significant reduction at $\Tc$ and exhibit 
	$1/T_1 \propto T$ well below $\Tc$. \cite{Sr2RuO4}
	This is the reason why we conclude the $1/T_1 \propto T$ 
	is not due to the impurity
	scattering but due to the odd-frequency gap.

\section{Coexistence of AF and SC order}
\begin{figure}[btp]
\begin{center}\leavevmode
	\includegraphics[width=0.7\linewidth]{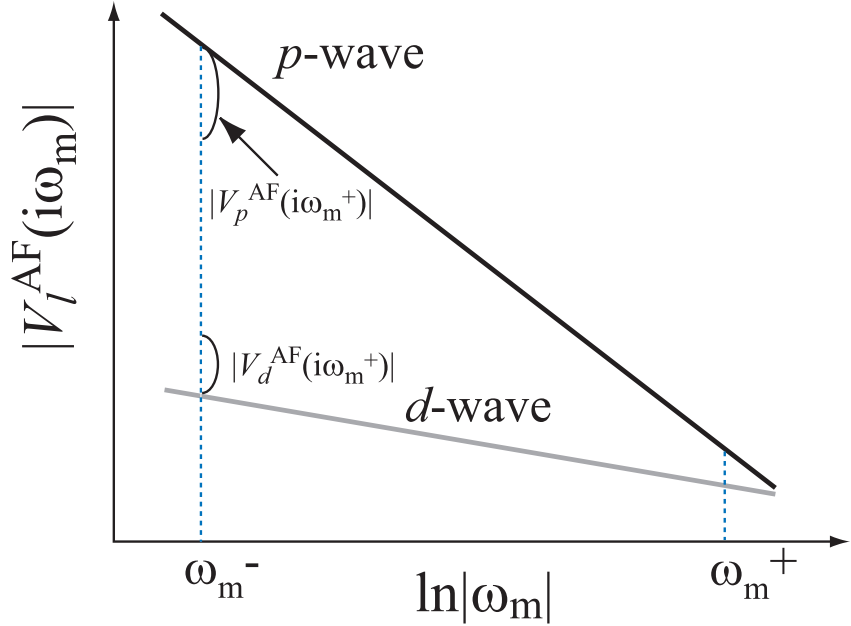}
	\caption{Illustration of the frequency dependence of $V_l (\textrm{i}
	\omega_m ) $in the AF backgrounds,
	where $\omega_m^\pm \equiv \varepsilon_n \pm \varepsilon_n '$. 
	The $p$-wave channel dominates the $d$-wave channel
	because $v_p > v_d $ and the unsaturated behavior of 
	$V_l (\textrm{i}\omega_m ) $.
}\label{VlAF}
\end{center}
\end{figure}

	In the previous sections, we discussed the emergence of the gapless
	$p$-wave singlet pairing prevailing the $d$-wave singlet
	pairing in the paramagnetic (PM) backgrounds.
	The experimental results \cite{Ishida,YKawasaki1,SO5,YKawasaki2,
	Mito,SKawasaki,Kitaoka1,Fisher} suggest that the gapless
	superconductivity is realized in the AF backgrounds
	rather than in the PM backgrounds. 
	(In the SC phase very close to the phase boundary, the present 
	theory predicts that the gapless $p$-wave singlet superconductivity
	is realized, but the detailed experimental results in this region
	have not been obtained yet.)
	Here we discuss that the gapless $p$-wave singlet can also be realized
	in rather wide region in AF+SC phase.
	Since the transverse spin susceptibility $\chi _{\perp}$
	would be much dominant than the longitudinal one $\chi _{\parallel }$,
	even if the damping effect would have been taken into account,
	the pairing interaction in the AF backgrounds may given as follows:
\begin{eqnarray}
	V(\textbf{q}, \textrm{i}\omega_m ) = 
	g^2 \chi_{\perp}(\textbf{q}, \textrm{i}\omega_m )
	\equiv \frac{g^2 N_F}{S^2 \hat{\textbf{q}}^2 +|\omega_m |^2},
	\label{AFint}
\end{eqnarray}
	where $S$ corresponds to the spin-wave velocity.	
	This interaction is similar to that just at QCP. 
	Namely, the frequency dependence
	of $V_l ^{\textrm{AF}}(\textrm{i} \omega _m )$ can be approximated as 
\begin{eqnarray}
	V_l ^{\textrm{AF}}(\textrm{i} \omega _m )\simeq v_l \ln \frac{\omega_0 '}
	{|\omega_m |^2}.
\end{eqnarray}
	Comparing this with the expression (\ref{appVl}), 
	the interaction $V_l ^{\textrm{AF}}$ of the AF side
	can be obtain by the transformation 
	$V_l ^{\textrm{AF}} (\textrm{i} \omega _m )\to 2V_l ^{\textrm{PM}}
	(\textrm{i} \omega _m ; \tilde{\eta} =0)$.
	Figure \ref{VlAF} illustrates this situation.
	There is no saturation in contrast to $V_l ^{\textrm{PM}}$ 
	in the PM side (see Fig.
	\ref{Vl} ), which stimulates the emergence of $p$-wave singlet pairing.
	How the $p$-wave states dominate the $d$-wave one 
	can be understood in the similar scenario discussed in \S 3,
	but this time we present much more generalized and intuitive 
	picture about how the $p$-wave dominates the $d$-wave.
	Each $V_l ^{\textrm{AF}}$ can be expressed in the form
\begin{eqnarray}
	V_p ^{\textrm{AF}}(\textrm{i} \omega _m ) &\simeq &
	v_p \Biggl( \ln \frac{1}{\omega_m^-} -\ln \frac{1}{\omega_m^+}
	\Biggr) \\
	V_d ^{\textrm{AF}}(\textrm{i} \omega _m ) &\simeq &
	v_d \Biggl( \ln \frac{1}{\omega_m^-} +\ln \frac{1}{\omega_m^+}
	\Biggr),
\end{eqnarray}
	where $\omega_m^{\pm } \equiv \varepsilon_n \pm \varepsilon_n '$.
	From these expressions, we can easily find that 
	$V_d ^{\textrm{AF}}>V_p ^{\textrm{AF}}$, in general, for $v_d \sim v_p$
	because the first term and the second term are added for
	$d$-wave, while they offset each other for $p$-wave.
	In the case $v_p$ is moderately larger than 
	$v_d $ as displayed in Fig. \ref{VlAF}, 
	however, the first term of $V_p ^{\textrm{AF}}$ is much larger
	than $V_d ^{\textrm{AF}}$, so that
	the $p$-wave singlet pairing dominates the $d$-wave one.
	This situation would be realized easier in the AF side
	rather than that in the PM side because of the factor 2 and 
	the unsaturation behavior of $V_l ^{\textrm{AF}}$.
	Judging from above, no matter how the detail is, 
	the $p$-wave pairing prevails $d$-wave one when the frequency
	dependence of the pairing interaction
	takes the form as shown in Fig. \ref{VlAF}.
\begin{figure}[btp]
\begin{center}\leavevmode
	\includegraphics[width=0.5\linewidth]{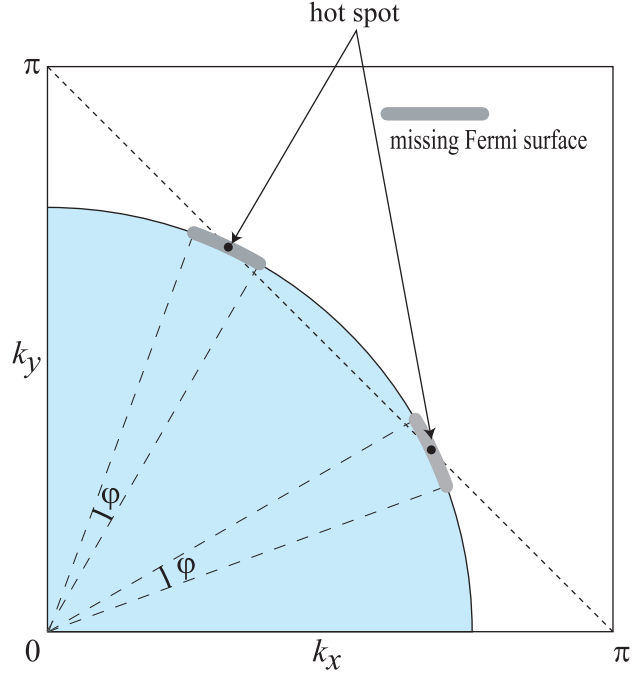}
	\caption{Missing Fermi surface due to the AF energy gap and 
	the definition of $\varphi$.
}\label{hotspot}
\end{center}
\end{figure}

	Below $T_{\textrm{N}}$, however, 
	the energy gap is formed around the hot spot
	(Fig. \ref{hotspot}) and the missing part of the Fermi surface
	spreads out as the AF ordering is developed.
	It leads to the remarkable reduction of $\Tc$ because
	the dominant pair scattering is suppressed due to the existence of
	the AF gap.
	To simulate this situation, we parameterize the 
	effect of the AF order as the 
	angle $\varphi$ of the missing part of the Fermi surface as displayed 
	in Fig. \ref{hotspot}.
	The growth of the AF gap, 
	spreading of $\varphi$, deprives the pairing interaction 
	of the most singular scattering with
	the AF ordering vector $\textrm{Q}$,
	and suppresses $V_l ^{\textrm{AF}}$.
	It is noted that this suppression is not the saturation effect displayed 
	in Fig. \ref{Vl}, but the reduction of $v_l$.
	This reduction of $v_l$ is remarkable for $v_p$, because
	the enhancement of $v_p$ is mainly due to the scattering 
	using the hot spots.

	We can also discuss the gapless nature as discussed in \S 4.
	In the AF backgrounds, the pairing interaction $V_l ^{\textrm{AF}}$
	is given by the same form as that in the PM backgrounds 
	with $v_l \to 2v_l$ in the limit $\eta \to 0$.
	Thus we reach the same conclution that the gap function in the AF side
	is always gapless:
\begin{eqnarray}
	\Delta _{\textrm{AF}}' (0) &=& 0, \\
	\Delta _{\textrm{AF}}'' (0) &=& 0. 
\end{eqnarray}

\begin{figure}[tbp]
\begin{center}\leavevmode
	\includegraphics[width=0.7\linewidth]{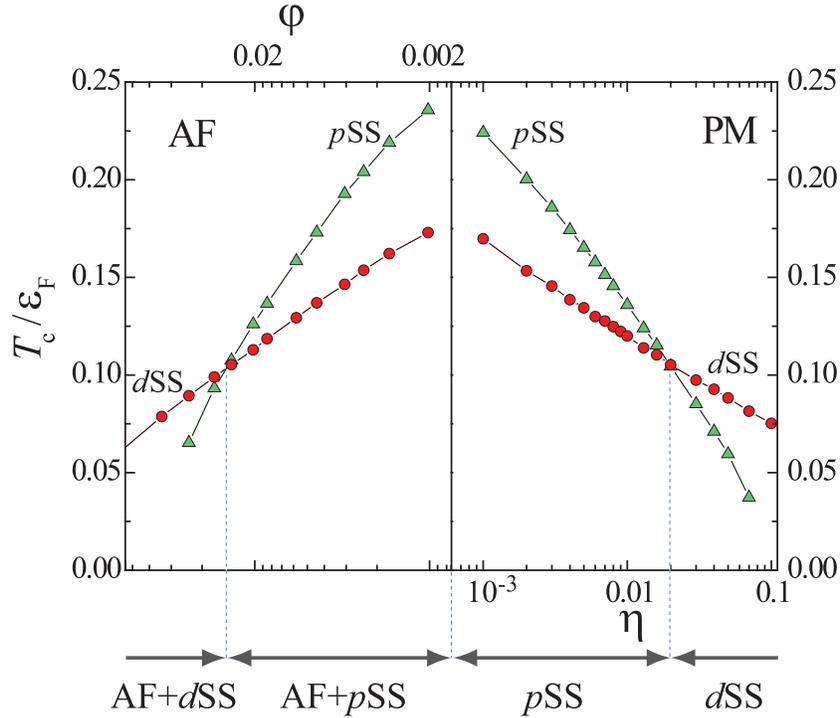}
	\caption{Transition temperatures near the QCP.
	AF+$d$SC: $d$-wave singlet pairing ($d$-S) with line nodes; 
	AF+$p$SC: gapless $p$-wave singlet ($p$-S);
	$p$SC: gapless $p$-S; $d$SC: $d$-S with line nodes.
}\label{etapsi}
\end{center}
\end{figure}

	Together with the $\Tc$'s in the PM phase,
	the calculated $\Tc$'s near the QCP are shown in Fig. \ref{etapsi}
	Here we set $S=1.2$ and the other parameter $g^2 N_F$ 
	and $k_F$ to be the same as those in the PM backgrounds.
\begin{figure}[tbp]
\begin{center}\leavevmode
	\includegraphics[width=0.7\linewidth]{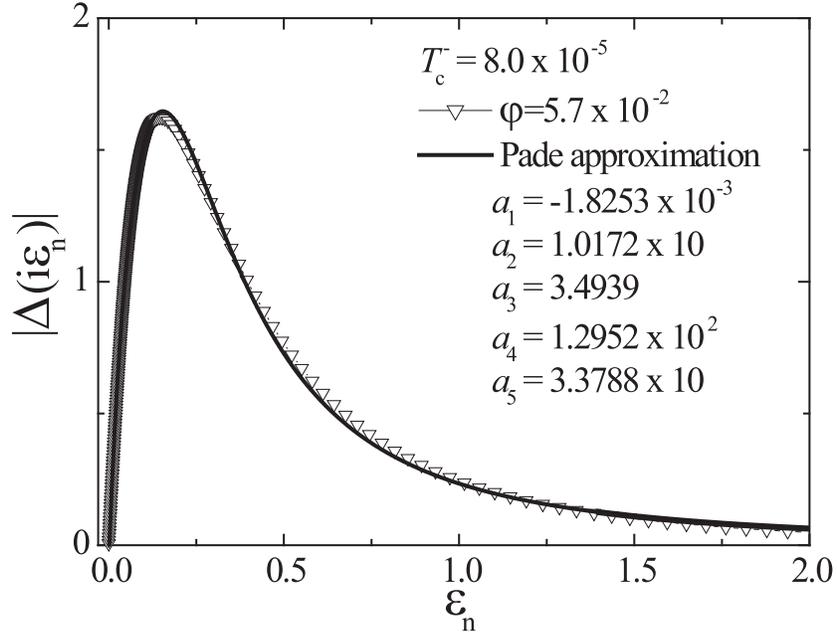}
	\caption{Frequency dependence of the gap function of the $p$-wave
	singlet pairing in the AF background. 
	The solid line indicates the results of Pad\'e approximation.
}\label{AFgap}
\end{center}
\end{figure}
	In the PM region, the distance from the QCP is 
	parameterized by $\eta$, and in the AF region, that is parameterized
	by $\varphi$.
	Indeed, the gap in AF backgrounds, shown in Fig. \ref{AFgap},
	has a gapless-like structure, and the parameters of Pad\'e approximation
	are obtained as $a_1 =-3.3415\times 10^{-3}, a_2 =2.5624 \times 10, 
	a_3 =-7.7475 \times 10, a_4 = 3.3734 \times 10^2, a_5 =4.8612 \times 10$.

	The gapless odd-frequency $p$-wave singlet pairing is realized only near
	the QCP, and $d$-wave singlet pairing prevails
	$p$-wave singlet pairing in the PM region away from the QCP.
	According to this phase diagram, the observed $1/T_1$ may be explained
	as follows.
	Near the QCP, i.e., AF+$p$SC and $p$SC region displayed 
	in Fig. \ref{etapsi},
	$1/T_1$ does not show any significant reduction due to 
	the gapless SC state.
	Away from the QCP (AF+$d$SC and $d$SC region),
	$1/T_1 \propto T^3$ behavior is observed.
	This phase diagram agrees qualitatively with several experiments
	in CeCu$_2$Si$_2$ and CeRhIn$_5$.
	Polycrystalline sample of Ce$_{0.99}$Cu$_{2.02}$Si$_2$ shows gapless
	behavior at ambient pressure, and $1/T_1 \propto T^3$ behavior
	at pressures $P \simg 0.1$GPa.\cite{Ishida,YKawasaki1}
	This property can be understood that Ce$_{0.99}$Cu$_{2.02}$Si$_2$
	at $P=0$GPa is located in $p$SC phase and at $P \simg 0.1$GPa, $d$SC
	phase.
	CeCu$_2$(Si$_{0.99}$Ge$_{0.01}$)$_2$ compounds shows the AF ordering
	at $T<0.75$K, and below 0.5K, the gapless superconductivity,
	coexisting with the AF order, at ambient pressure.
	This compounds also exhibits the line-node gap under the pressure
	$P=0.85$GPa.\cite{YKawasaki2}
	Therefore, this compounds would be located in AF+$p$SC phase at 
	ambient pressure and in $d$SC phase at $P=0.85$GPa.
	The existence of $p$SC phase in this sample cannot be recognized
	from the present experimental data.
	Similarly, in CeRhIn$_5$ the pressure region $1.6\siml P \siml 1.75$GPa
	would correspond to AF+$p$SC and $P>2.1$GPa 
	would correspond to $d$SC.\cite{Mito,SKawasaki,Kitaoka1}
	For each case, the AF+$d$SC phase, where the $d$-wave singlet
	SC states and the AF states coexist, have not been observed yet.

\section{Conclusion}
	In the present paper, we have shown that the $p$-wave singlet
	superconductivity with the gap function which is odd in both
	momentum and frequency prevails the $d$-wave singlet
	superconductivity, at the AF-QCP and in the AF side of AF-QCP.
	This odd-frequency $p$-wave singlet pairing is stabilized in
	the AF states, and is able to coexist with the AF order.
	The characteristic properties of this $p$-wave 
	singlet superconductivity is that
	i) there is no gap in the quasiparticle spectrum;
	ii) $d$-wave pairing arises apart from the AF-QCP even in the AF state;
	iii) In some parameter region, $p$-wave pairing is realized only 
	in the window of temperatures $\Tc^- <T<\Tc^+$.
	The first and the second property would explain the experiments.
	Namely, the $1/T_1 \propto T$ behavior is observed in the AF+SC region or 
	in the boundary of CeCu$_2$Si$_2$ or CeRhIn$_5$, 
	and $1/T_1 \propto T^3 $ behavior 
	in the PM side away from the boundary.
	From the second one, we can predict that
	the $d$-wave pairing will be observed again away from the AF-QCP
	in the AF states.

	The condition of the emergence of the $p$-wave singlet pairing
	is that the FS is not nested, so that the hot points of surface
	connected by the AF wave vector	$\textbf{Q}=(\pi, \pi )$ are isolated,
	while the AF order is induced by the exchange interaction between
	localized component of spin degrees of freedom.
	Although the structure of the Fermi surface of CeCu$_2$Si$_2$
	\cite{Harima} and CeRhIn$_5$\cite{Settai}
	is much more complicated than the present model,
	they seem to satisfy the condition from inspection of the shape of 
	the Fermi surface obtained by band structure calculations.
	We thus believe that the present theory would distill the fundamental
	picture of both CeCu$_2$Si$_2$ and CeRhIn$_5$, and obtain the qualitative
	understanding of them.
	In order to get futher understanding, extensive calculation
	considering the momentum-dipendence and the practical dispersion
	of the quasiparticle are now in progress.

\section*{Acknowledgements}
	Authors would like to thank S. Kawasaki, G.-q.\ Zheng and Y. Kitaoka 
	for discussion of experimental results.
	One of the authors (Y. F) is supported by Research Fellowships of the 
	Japan Society for the Promotion of Science for Young Scientists.
	This work was supported by a Grant-in-Aid for COE Research (10CE2004)
	from Monbu-Kagaku-sho.


\end{document}